\documentclass[usenatbib,iop,apj,numberedappendix]{aeb_emulateapj_2010}
\usepackage{natbib}
\usepackage{amsmath}
\usepackage{amssymb}
\usepackage{amsbsy}
\usepackage{xspace}

\usepackage[normalem]{ulem}
\usepackage[usenames]{color}

\def\eV{{\rm eV}} 
\def\GeV{{\rm G}\eV} 

\newcommand{\rmn}{\mathrm}

\newcommand{\kvec}{{\bf k}}

\def\G{{\rm G}}

\def\Fermi{{\em Fermi}\xspace}

\begin{document}

\title{The Effect of Nonlinear Landau Damping on Ultrarelativistic Beam Plasma Instabilities} 

\author{
Philip Chang\altaffilmark{1},
Avery E.~Broderick\altaffilmark{2,3},
Christoph Pfrommer\altaffilmark{4},
Ewald Puchwein\altaffilmark{5},
Astrid Lamberts\altaffilmark{1},
Mohamad Shalaby\altaffilmark{2,3,6}
}
\altaffiltext{1}{Department of Physics, University of Wisconsin-Milwaukee, 1900 E. Kenwood Boulevard, Milwaukee, WI 53211, USA}
\altaffiltext{2}{Perimeter Institute for Theoretical Physics, 31 Caroline Street North, Waterloo, ON, N2L 2Y5, Canada}
\altaffiltext{3}{Department of Physics and Astronomy, University of Waterloo, 200 University Avenue West, Waterloo, ON, N2L 3G1, Canada}
\altaffiltext{4}{Heidelberg Institute for Theoretical Studies, Schloss-Wolfsbrunnenweg 35, D-69118 Heidelberg, Germany}
\altaffiltext{5}{Institute of Astronomy and Kavli Institute for Cosmology, University of Cambridge, Madingley Road, Cambridge CB3 0HA, UK}
\altaffiltext{6}{Department of Physics, Faculty of Science, Cairo University, Giza 12613, Egypt.}
\keywords{
BL Lacertae objects: general -- gamma rays: general -- plasmas -- instabilities -- magnetic fields
}

\begin{abstract}
Very-high energy gamma-rays from extragalactic sources pair-produce off of the extragalactic background light, yielding an electron-positron pair beam.  
This pair beam is unstable to various plasma instabilities, especially the ``oblique'' instability, which can be the dominant cooling mechanism for the beam.  However, recently, it has been claimed that nonlinear Landau damping  renders it physically irrelevant by reducing the effective damping rate to a low level. Here, we show with numerical calculations that the effective damping rate is $8\times 10^{-4}$ of the growth rate of the linear instability, which is sufficient for the ``oblique'' instability to be the dominant cooling mechanism of these pair beams.  In particular, we show that previous estimates of this rate ignored the exponential cutoff in the scattering amplitude at large wavenumber and assumed that the damping of scattered waves entirely depends on collisions, ignoring collisionless processes. We find that the total wave energy eventually grows to approximate equipartition with the beam by increasingly depositing energy into long wavelength modes.  As we have not included the effect of nonlinear wave-wave interactions on these long wavelength modes, this scenario represents the ``worst-case'' scenario for the oblique instability.  As it continues to drain energy from the beam at a faster rate than other processes, we conclude that the ``oblique'' 
instability is sufficiently strong to make it the physically dominant cooling mechanism for high-energy pair beams in the intergalactic medium.  
\end{abstract}

\section{Introduction}\label{sec:introduction}

The very-high energy gamma-ray (VHEGR, $E\ge100\,\GeV$) extragalactic sky is 
dominated by a subset of blazars, which we refer to as TeV blazars.  
These extragalactic VHEGR emitters produce TeV photons that are greatly attenuated via annihilation upon soft photons in the extragalactic background light
(EBL) and produce pairs \citep[see, e.g.,][]{Goul-Schr:67,Sala-Stec:98,Nero-Semi:09}.

It has been assumed that these ultrarelativistic pairs produced by VHEGR annihilation 
lose energy exclusively through inverse-Compton scattering off of the cosmic microwave
background (CMB),   transferring the energy of the original VHEGR to gamma-rays with
energies $\lesssim100\,\GeV$.  As the gamma-rays are in the LAT bands of \Fermi, this is an important target for \Fermi observations.

The absence of observed secondary IC emission leads a number of
authors to argue that this \textit{lack} of observed emission places lower bounds upon the
intergalactic magnetic field
\citep[IGMF; see, e.g.,][]{Nero-Vovk:10,Tave_etal:10a,Tave_etal:10b,Derm_etal:10,Tayl-Vovk-Nero:11,Taka_etal:11,Dola_etal:11} with 
typical numbers ranging from $10^{-19}\,\G$ to $10^{-15}\,\G$.  In addition, the persistent belief in this IC emission has led other workers to argue that based on the IC contribution to Fermi extragalactic gamma-ray background (EGRB), the comoving number
density of gamma-ray bright blazars must grow slowly with increasing redshift, if
at all \citep{Knei-Mann:08,Vent:10,Abazajian:2011,Inoue:2012}, implying that these TeV blazars cosmologically evolve qualitatively differently compared to
other active galactic nuclei (AGNs).

These two conclusions depend on IC cooling
dominating the evolution of the ultra-relativistic pairs.  However, it was recently found that plasma instabilities driven by the
ultrarelativistic pair beams  likely is the dominant cooling mechanism
\citep{paperI,Schlickeiser+12,Schlickeiser+13}, depositing this energy
as heat in the intergalactic medium
\citep{paperII,paperIII}.  Therefore, the lack of an observed
 IC halo emission from TeV blazars does not imply the 
existence of the IGMF as previous groups have argued
\citep{paperI,Schlickeiser+12,Schlickeiser+13}.

This excess heating of the IGM may resolve a variety
of cosmological puzzles, including naturally explaining
  anomalies in the statistics of the high-redshift Ly$\alpha$ forest
\citep{paperIV} and potentially explaining a number of the X-ray
  properties of groups and clusters and anomalies in galaxy formation on
  the scale of dwarfs \citep{paperIII,Lu+2013}.  We
  have recently shown that \textit{if} the IC halos are ignored, it is possible
  to quantitatively reproduce the redshift and flux distributions
  of nearby hard gamma-ray blazars and the extragalactic gamma-ray background
  spectrum above 3~GeV simultaneously with a unified model of AGN evolution
  \citep{Broderick+2013,Broderick+2013b}.  All of these empirical successes provide
  circumstantial evidence for the presence of virulent plasma beam
  instabilities. 

Recently, \citet[][hereafter ME13]{Miniati+12} argued that these instabilities are physically irrelevant for the cooling of these pair beams. First, the ``oblique'' instability would saturate at a very low level due to nonlinear Landau damping (NLD).  
ME13 argue that this process occurs when  $3\times 10^{-6}$ of the electron-positron beam energy is contained within the waves and significantly limiting the instability cooling rate. 
Second, inhomogeneities in the IGM prevent this linear instability from even growing by shifting the unstable waves out of the region of resonance. The second point relates to the properties of the linear instability in the kinetic regime, which has been shown by \citet{Schlickeiser+13} to be physically relevant, contrary to ME13's claims.  We  will address this second objection of ME13 in future work.  

In this paper, we will consider the effects of NLD. We begin by discussing the physics in \S \ref{sec:physics}. Using a numerical calculation, we derive the saturation level of the ``oblique'' instability in the non-linear regime in \S \ref{sec:averaged}.  We  discuss why our results differ from those presented by ME13 in \S \ref{sec:comparison} and present the implications in \S \ref{sec:implications}. Finally, we close with a summary of results and pathway to future work in \S \ref{sec:conclusions}.

\section{The Physics of Nonlinear Landau Damping}\label{sec:physics}

We consider an unstable wave that is driven by a beam of electrons and positrons in a background plasma of electrons and protons (or other ions).  As the unstable wave grows in amplitude, it becomes subject to nonlinear wave-particle and wave-wave interactions.  In nonlinear particle-wave interactions, the most important interaction is induced scattering by thermal ions \citep{Kaplan+68,Smith+71,Breizman+72,Melrose86}, which is also referred to as NLD \citep{Melrose86}. 

Wave-particle interactions induce the transformation of one plasma wave, characterized by a frequency and wavevector $(\omega,\kvec)$, into another $(\omega',\kvec')$ via nonlinear scattering on the particles that constitute the plasma.
The kinetic equation for these waves in the presence of wave-particle interactions is \citep{Kaplan+68,Breizman+72}
\begin{equation}\label{eq:nonlinear}
 \frac{d W_{\kvec}}{d t} = 2\Gamma_{\kvec} W_{\kvec} - \frac {W_{\kvec}\omega_p}{8 (2\pi)^{5/2} n_e m_e v_e^2}\int \frac{(\kvec\cdot\kvec')^2}{k^2 k'^2}\phi(\kvec,\kvec')W_{\kvec'}d\kvec'
\end{equation}
where $W_{\kvec}$ is the spectral energy density of the waves, normalized such that the total energy density is given by
\begin{equation}
  \label{eq:norm}
  W = \frac {1}{(2\pi)^3} \int W_{\kvec} d\kvec\,,
\end{equation}
$\Gamma_{\kvec} = \Gamma_{\rm gr}(\kvec) + \Gamma_{\rm LD}(\kvec)$ is the sum of the unstable wave growth rate, $\Gamma_{\rm gr}(\kvec)$,  and the linear Landau damping rate, $\Gamma_{\rm LD}(\kvec)$, $\omega_p\equiv\sqrt{4\pi n_e e^2/m_e}$ is the electron plasma frequency of the background plasma, given in terms of the proper electron density ($n_e$) and rest mass ($m_e$), and $v_e$ and $v_i$ are the electron and ion thermal velocities, respectively.  We also note that wave-particle interactions also convert these electrostatic waves into electromagnetic waves as noted in \citet{Kaplan+68}.  However, we ignore electromagnetic modes in this work to focus on the the electrostatic waves.  Including these modes as well other nonlinear processes is the subject of future work.

The growth rate is given by the oblique growth rate, which is
\begin{equation}\label{eq:growth rate}
 \Gamma_{\rm gr}(\kvec) \equiv\frac{1}{\tau_{\rm gr}} \approx 0.4 \frac{n_b}{n_e} \gamma_b \omega_p \Theta(1 - v_{\rm ph}(k)/c), 
\end{equation}
where $n_b$ is the beam density and $\gamma_b$ is the Lorentz factor of the ultrarelativistic beam, $\Theta$ is the Heaviside function, and $v_{\rm ph}(k) = \omega/k$ is the phase speed of the Langmuir wave.  Equation (\ref{eq:growth rate}) was first found by \citet{Bret+10} by fitting the maximum growth rate in the kinetic regime. 
We have confirmed this growth rate in the electrostatic approximation for the kinetic regime (Broderick et al., in prep., see also \citealt{Schlickeiser+13}).  More importantly, this result holds true for a large range in $\kvec$.  The reason for this is that electrostatic waves oscillate at $\omega_p$ almost independently of $\kvec$ and their phase speed along the z-axis (arbitrarily defined) is given by $v_{\rm ph} = \omega/k\cos\theta$, where $\theta$ is the angle between the direction in question and the wave vector.  Hence for any $k \geq \omega_p/c$, there exist some $\cos\theta$ such that the $v_{\rm ph} \approx c$ and hence these waves are in resonance with a relativistic beam.  In other words, the oblique instability grows for any $k\geq\omega_p/c$, but the angle between the most unstable wavevector and beam varies.

Damping in the linear regime is given by linear Landau damping, whose rate is
\begin{equation}\label{eq:linear_landau}
\Gamma_{\rm LD}(\kvec) = -\left(\frac{\pi}8\right)^{1/2} \omega_p \left(\frac{v_{\rm ph}}{v_e}\right)^3\exp\left(-\frac{v_{\rm ph}^2}{2 v_e^2}\right).
\end{equation}

The overlap integral $\phi$ is given by \citep{Kaplan+68}
\begin{equation}\label{eq:induced rate}
 \phi(\kvec,\kvec') = \frac{3 v_e^2\left(k^2- k'^2\right)}{4\omega_p|\kvec - \kvec'| v_i}\exp\left[-2\left(\frac{3 v_e^2\left(k^2- k'^2\right)}{4\omega_p|\kvec - \kvec'| v_i}\right)^2\right].
\end{equation}
An important feature of equation (\ref{eq:induced rate}) is the dependence on ${k'}^2-k^2$, which sets the sign of $\phi(k,k')$.  Scattering of a wave with
wavevector $\kvec$ into another wave with wavevector $\kvec'$ can only proceed if $\phi(k,k')>0$, i.e., the wave energy in the $\kvec$ wave is damped, while the 
$\kvec'$ wave energy grows.  This requires that $k'<k$, i.e., the scattered wave has a longer wavelength than the incident wave. The demand that induced scattering drives waves 
to longer wavelength arises from the transfer of some momentum from the incident wave into the polarization clouds surrounding the ions \citep{Smith+71}.

\section{Numerical Studies}\label{sec:averaged}

We now solve equation (\ref{eq:nonlinear}) numerically assuming that linear growth and nonlinear damping via NLD are the two mechanisms that control the initial evolution of Langmuir waves.  However, because $\kvec$ is three-dimensional, we adopt the simplifying assumption that the Langmuir waves are isotropic, i.e., $W_{\kvec} = W_{k}$. 
This simplifying assumption is reasonable as long as the induced
  scattering processes are sufficiently rapid that it isotropizes the
  waves\footnote{Further support for this approximation emerges if
      the number of TeV blazars that contribute to the heating of a given patch of the intergalactic medium exceeds 100.} \citep{Kaplan+68}.
Equation (\ref{eq:nonlinear}) reduces then to\footnote{Equation (\ref{eq:induced rate}) can also be simplified if we set ${(\kvec_1 \cdot \kvec_2)^2}/{k_1^2 k_2^2}$ to the angle-averaged value of $1/3$ as done by \citet{Kaplan+68}.  We computed this integral both ways and found little difference in the saturation amplitude.}
\begin{eqnarray}
\frac{d W_{k}}{d t} &=& 2\Gamma_{k} W_{k} - \frac {W_{k}\omega_p}{8 (2\pi)^{3/2} n_e m_e v_e^2}\nonumber\\ &&\times \int k'^2\cos^2\theta\,\phi(\kvec,\kvec')W_{k'} dk' d\cos\theta
\label{eq:nonlinear reduced}
\end{eqnarray}
where $\theta$ is the angle between $\kvec$ and $\kvec'$.  Without loss of generality, we have fixed $\kvec$ along the z-axis. Here $\phi(\kvec,\kvec')$ can be simplified to 
\begin{eqnarray}
 \phi(\kvec,\kvec') &=& \frac{3 v_e^2\left(k^2- k'^2\right)}{4\omega_p v_i\sqrt{k^2 + k'^2 - 2kk'\cos\theta}}\nonumber\\
 &&\times\exp\left[-2\left(\frac{3 v_e^2\left(k^2- k'^2\right)}{4\omega_p v_i\sqrt{k^2 + k'^2 - 2kk'\cos\theta}}\right)^2\right]. 
\end{eqnarray}

We calculate equation (\ref{eq:nonlinear reduced}) numerically for
$N_{\rm modes} = 300$ logarithmically spaced modes from $k = 10^{-6}
\omega_p/c$ to $10^3 \omega_p/c$ and have verified this calculation
using $N_{\rm modes} = 1000$.  The lower limit of $k_{\rm min}$ was chosen to fulfill
the requirement $k_{\rm min} \ll \omega_p/c$.  The upper limit of $k_{\rm max} = 10^3
\omega_p/c$ was set because it is significantly larger than the estimated $k$ where linear Landau
damping would suppress the instability.
We have found that our
  calculations are not affected by extending the upper and lower
  limits on k.  We also set the initial $k^3W_k$ to a small value of the initial beam energy, i.e., $10^{-15}$ and confirmed that our results are independent of this initial value.

\begin{figure}
 \plotone{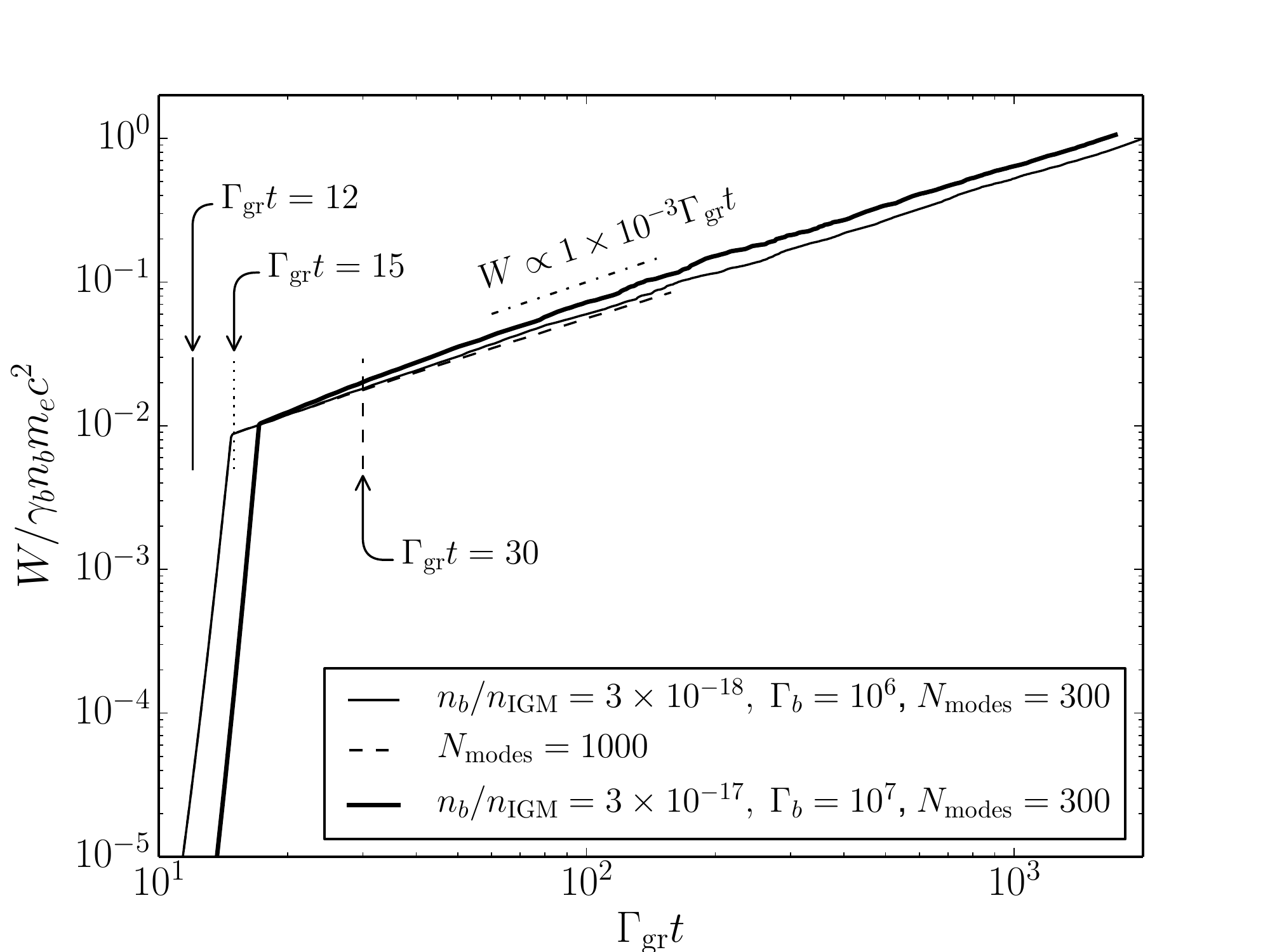}
 \caption{Wave energy, $W$, in units of the beam energy, $n_b \gamma_b m_e c^2$, as a function of growth e-folding times $\Gamma_{\rm gr} t$. The dot-dashed curve is a line defined by $W \propto 10^{-3}\Gamma_{\rm gr} t$.  \label{fig:saturation}}
\end{figure}

In Figure \ref{fig:saturation}, we plot the wave energy $W$ in units of the initial beam energy density $n_b \gamma_b m_e c^2$ as a function of growth e-folding times $\Gamma_{\rm gr} t$ for a $1$ TeV beam and a $10$ TeV beam, where the beam density is what is expected at $z=0$ for a TeV blazar with equivalent isotropic luminosity $E L_E=10^{45}~\rmn{erg~s}^{-1}$.  The wave energy grows exponentially up to a time $\Gamma_{\rm gr} t \approx 15$, where exponential growth ends and transitions to a slow linear growth in $W$.  The wave energy, $W$, is equal to the energy of the initial beam when $\Gamma_{\rm gr} t = 2000$. Therefore, in the absence of a significant back-reaction, the beam experiences one e-folding reduction in energy at $\Gamma_{\rm gr} t \approx 1300$, giving a damping rate of
\begin{equation}\label{eq:average_damping_rate}
\Gamma_{\rm NLD} \approx 8\times 10^{-4} \Gamma_{\rm gr}.
\end{equation}

\begin{figure}
 \plotone{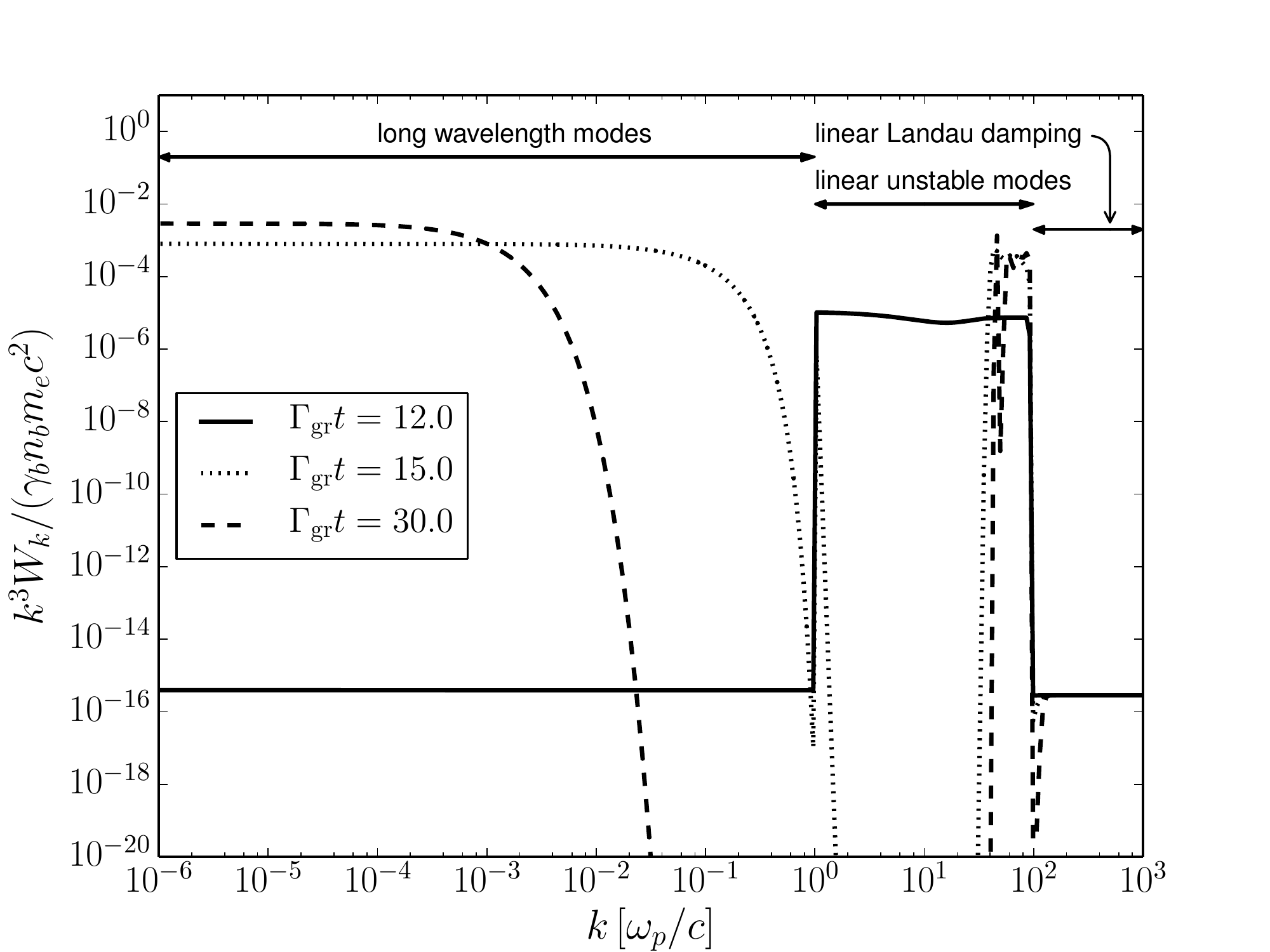}
  \caption{Wave energy, $k^3 W_k$, in units of the beam energy, $n_b \gamma_b m_e c^2$, as a function of wavevector $k$ for three different times $\Gamma_{\rm gr} t = $12 (solid line), 15 (dotted line), and 30 (dashed line).  \label{fig:modes}}
\end{figure}

In Figure \ref{fig:modes} we plot the wave energy as a function of wavevector, $k$, for three different times, $\Gamma_{\rm gr} t =$ 12 (solid line), 15 (dotted line), and 30 (dashed line).  These times are also marked in Figure \ref{fig:saturation} by vertical lines of the same type as in Figure \ref{fig:modes}.  For $\Gamma_{\rm gr} t = $ 12 (solid line), NLD is not important and the instability grows for all $k \geq \omega_p/c$. As the unstable waves grow, the effect of NLD begins to become important and long wavelength modes $k < \omega_p/c$ begin to grow at the expense of short wavelength modes.  This is clearly seen for $\Gamma_{\rm gr} t = $ 15 (dotted line) and 30 (dashed line).  However, the largest wavevector modes are not suppressed by the effect of NLD and remain at a level of $\approx 10^{-3}$ of the beam energy density.  These large wavevector modes survive for the duration of the calculation and continually pump energy into long wavelength modes.  

To understand the origin of the survival of these large wavevector modes and the development of the empty region between $k\approx 10^{-2} \omega_p/c$ and $\sim 20 \omega_p/c$, it is helpful to return to the coupling term (\ref{eq:induced rate}).  Because the argument in the exponent is dominated by $k'\approx k$, modes that are closely spaced to each other are strongly scattered. When the difference is large, i.e., $k^2 - k'^2$ is large, the scattering is exponentially suppressed.  Therefore, NLD is strongest on modes near each other.   As a result, as the long wavelength modes grow, they quickly sap energy from the nearest modes, leaving the short wavelength modes untouched due to the exponential suppression. This suppression becomes important where the exponent is order unity or 
\begin{equation}
 \frac {v_e^2\left(k^2- k'^2\right)}{\omega_p|\kvec - \kvec'| v_i} \approx \frac{v_e^2 k}{\omega_p v_i} \gtrsim 1
\end{equation}
This gives the condition 
\begin{equation}
 k \gtrsim \sqrt{\frac{m_e}{m_i}} \frac{\omega_p}{c} \frac{c}{v_e} \approx 20 \left(\frac {T} {10^4\,{\rm K}}\right)^{-1/2}\frac{\omega_p}{c}, 
\end{equation}
which is of order where the suppression occurs (see Figure \ref{fig:modes}).

\section{Relationship to Previous Work}\label{sec:comparison}

\subsection{Comparison with \citet{Miniati+12}}
Our result differs from ME13, who found that the effect of NLD on the ``oblique'' instability is to drive the saturation of the excited Langmuir waves to a physically irrelevant amplitude.  
ME13 estimates the excited wave energy to be $W/(n_b \gamma m_e c^2) \approx 3\times 10^{-6}$, whereas our numerical calculation finds a value that is over two orders of magnitude larger.  There are two crucial differences.  First, ME13 based their estimate on the order-of-magnitude calculation which ignores the effect of the suppression of NLD at large wavenumber.  Second, ME13 assumes that the damping of these scattered waves is completely due to collisions, which are extraordinarily slow.  In this case, the wave energy of the excited waves is reduced to 
${W_r}/{\gamma_b n_b m_e c^2} = {\Gamma_{\rm NLD}}/{\Gamma_{\rm c}},$
where $W_r$ is the wave energy of resonant waves, $\Gamma_{\rm c}$ is the electron collision rate and is typically given by electron-ion collisions, and $\Gamma_{\rm NLD}$ is the maximum growth rate of the resonant waves that are unstable to growth due to NLD of the linearly unstable mode.  

The choice of collisional damping  likely underestimates the true damping rate.  In particular, collisionless processes like wave-wave scattering and wave-particle interaction are likely to produce a damping rate for these scattered waves that far exceeds collisional damping.  Indeed, collisionless simulations \citep{Davidson72} have shown that nonlinear wave-wave interactions can lead to particle heating and wave damping in the absence of collisions.  

Here we presume that the damping rate of the scattered waves are sufficiently rapid to give
\begin{equation}
\frac{W_r}{\gamma_b n_b m_e c^2} = \frac{\Gamma_{\rm NLD}}{\Gamma_{\rm gr}} \approx 8\times 10^{-4}.
\end{equation}
We are motivated to use this estimate by comparison with the work of \citet{Ziebell+08a, Ziebell+08b} who found while NLD effects dominate the scattering of Langmuir waves, the rate of three-wave interactions is competitive with NLD and results in the quasi-isotropic heating of the electrons. The results of \citet{Ziebell+08a,Ziebell+08b} suggest that collisional processes are irrelevant to the damping of scattered waves.  Moreover, nonlinear wave-wave coupling will likely lead to a more equitable distribution of mode energies among the different wavevectors whereas the effect of NLD moves energy toward smaller wavevectors. Thus, inclusion of nonlinear wave-wave coupling will lead to a stronger damping rate for the oblique instability by countering the effect of NLD.  As a result, the calculation that we present here likely represents the worst case scenario for the ``oblique'' instability. 

\subsection{Comparison with \citet{Sironi+14}}
Recent numerical simulations have been brought to bear upon these dilute beam-plasma processes to study the physics of nonlinear saturation by \citet[][hereafter SG14]{Sironi+14}.  However, due both to their spectral resolution and constraints on the parameters simulated these appear to be unable to capture the physics of NLD.

The range of $k$ over which NLD redistributes energy is limited by the particular form of the overlap integral, eq. \ref{eq:induced rate}.  To estimate this range, we note that the overlap integral has a maximum width for $\kvec'$ antiparallel to $\kvec$, i.e., the backward scattering case; in this case, we take 
$\kvec' = -(k-\Delta k)\hat{\kvec}$ and apply this to equation (\ref{eq:induced rate}) to give 
\begin{equation}
 \phi(\Delta k) = \frac{3v_e^2\Delta k}{4\omega_p v_i} \exp\left[-2\left(\frac{3v_e^2\Delta k}{4\omega_p v_i}\right)^2\right],
\end{equation}
for $\Delta k/k \ll 1$.  This peaks at
${3v_e^2\Delta k}/{4\omega_p v_i} =1/2$, 
falling shortly thereafter.  Thus, to marginally resolve NLD requires a spectral resolutions in excess of
\begin{equation}\label{eq:Delta k}
  \frac{\Delta k}{\omega_p/c} = \frac{2 v_i c}{{3} v_e^2} 
= \frac{2}{{3}} \sqrt{\frac{m_e c^2}{kT_e} \frac{m_e}{m_p}\frac{T_i}{T_e}},
\end{equation}
SG14 found that the temperature of the background electrons approached relativistic temperatures (see Figure 3 of SG14) in their simulations and hence $\Delta k \approx 0.02\omega_p/c$, assuming $T_i=T_e$.  

The spectral resolution of a numerical computation is set by the simulated domain's physical size.  In SG14 the reported size is $128\,c/\omega_P$, implying a spectral resolution of $\Delta k_{\rm min} = (2\pi/128) \omega_p/c \approx 0.05 \omega_p/c$, larger than the minimum required to reolve the NLD.  Therefore, the calculations described in SG14 are unable to capture the impact of NLD even for the most optimistic case of backward scattering.  Moreover, it is likely that $T_i \ll T_e$ as the collisions needed to maintain this equilibrium are absent and electromagnetic interactions are inefficient because of the large mass ratio; a significantly lowered $T_i$ would make this disparity even more substantial.

In addition, the high temperatures reached in SG14's simulations suppresses the effectiveness of NLD.  To see this, we estimate the NLD term in equation~(\ref{eq:nonlinear reduced}) for the backward scattering case discussed above.  Again taking $\kvec' = -(k-\Delta k)\hat{\kvec}$ and integrating over $\Delta k$, we find equation~(\ref{eq:nonlinear reduced}) becomes
\begin{equation}
 \frac{d W_{k}}{d t} \approx 2\Gamma_k W_{k} - \frac {W_{k}\omega_p}{8 (2\pi)^{5/2} n_e m_e v_e^2} \frac{4\pi} 3 k^2\Delta k W_k, 
\end{equation}
where we have approximated the $\phi(\kvec,\kvec') \approx \phi(\Delta k) \approx 1/3$ over the interval $\Delta k$ given by equation (\ref{eq:Delta k}). Substituting $\Delta k$ by equation (\ref{eq:Delta k}) and $\Gamma_k$ by equation (\ref{eq:growth rate}), the ratio between the second and first terms, indicating the importance of NLD, is 
\begin{equation} 
 \Gamma_k^{-1}\frac {4\pi\omega_pk^2\Delta k W_k}{48 (2\pi)^{5/2} n_e m_e v_e^2} \approx 10^{-4} \frac{\omega_p}{kc} \frac{k^3 W_k}{\gamma_b n_b m_e c^2} \left(\frac{m_e c^2}{k_B T} \right)^{3/2}.
\end{equation}
For conditions relevant to SG14's simulations, $k^3 W_k/\gamma_b n_b m_e c^2 \approx 0.1$, $k_B T/m_e c^2 \approx 1$ and $\omega_p/kc \approx 1$, implying NLD is suppressed by five orders of magnitude compared to linear growth. 


It remains unclear if quasilinear relaxation plays an important role.  In the simulations performed by SG14 no more than 10\% of the energy of the original beam is drained by the oblique instability, which they attribute to quasilinear relaxation processes.  Assuming this to be the case, extrapolating to the parameters of intergalactic TeV-driven beams results in even more stringent limits on the efficiency of plasma instabilities.  However, the accuracy and applicability of these extrapolations, from $\gamma\approx10^2$ and $n_b/n_e\approx10^{-2}$ to $\gamma\approx10^6$--$10^7$ and $n_b/n_e\approx3\times10^{-18}$, is far from clear, and depends critically on the identification of the physical and potentially numerical causes of the instability saturation.  An exhaustive study of these effects, quasilinear relaxation and nonlinear wave-wave coupling in conjunction with NLD, is left for future work.


\section{Implications}
\label{sec:implications}

We now compare the damping rate given by equation (\ref{eq:average_damping_rate}) to the current (i.e., at $z=0$) damping rate due to inverse Compton scattering off cosmic microwave background (CMB) photons, which is given by \citet{paperI}:
\begin{equation}
\Gamma_{\rm IC} = \frac{4\sigma_T u_{\rm CMB}}{3m_e c} \gamma_b\approx 2.2\times 10^{-13} \left(\frac{E}{{\rm TeV}}\right)\left(\frac{1+z}2\right)^4\,{\rm s}^{-1}.\label{eq:ic_damping}
\end{equation}
This sets a maximum beam density (paper I), which we can use to get the effective maximum damping rate of the beam.
For $\Gamma_{\rm NLD} = 8\times 10^{-4} \Gamma_{\rm gr}$, we find
\begin{equation}\label{eq:max}
 \Gamma_{\rm NLD, max} \approx 6.9\times 10^{-12} \left(\frac{1+z}{2}\right)^{3\zeta-5.5}\left(\frac{EL_E}{10^{45}\,{\rm ergs\,s}^{-1}}\right) \left(\frac {E} {\rm TeV}\right)^2\,{\rm s}^{-1},
\end{equation}
where $\zeta$ is our parameterization of the extragalactic background light and is given by $\zeta = 4.5$ for $z < 1$ and 0 otherwise.  

However, the above calculation is not self-consistent as we have presumed that the beam density is limited only by inverse Compton scattering.  If instead we assume that the damping of the beam is driven by these plasma instabilities \citep{paperI} with the damping rate given by equation (\ref{eq:average_damping_rate}), we find a self-consistent
effective damping rate of
\begin{multline}
 \Gamma_{\rm NLD} \approx 10^{-12} \left(\frac{1+z}{2}\right)^{(6\zeta-3)/4}\left(\frac{EL_E}{10^{45}\,{\rm ergs\,s}^{-1}}\right)^{1/2}\left(\frac{E}{\rm TeV}\right)^{3/2}\,{\rm s}^{-1}\label{eq:nld_damping}
\end{multline}
Hence, we find that NLD while important does not appear to be sufficiently strong to prevent the oblique instability from dominating the cooling of the pair beam at energies $\gtrsim 0.8$ TeV.  This domination is not complete at $z=0$: a substantial fraction of the beam energy can now be lost to inverse Compton scattering, of order 46\% at 1 TeV and 20\% at 10 TeV.  We caution that this value may be significantly reduced if nonlinear wave-wave interactions reduce the effectiveness of NLD.

\begin{figure}
 \plotone{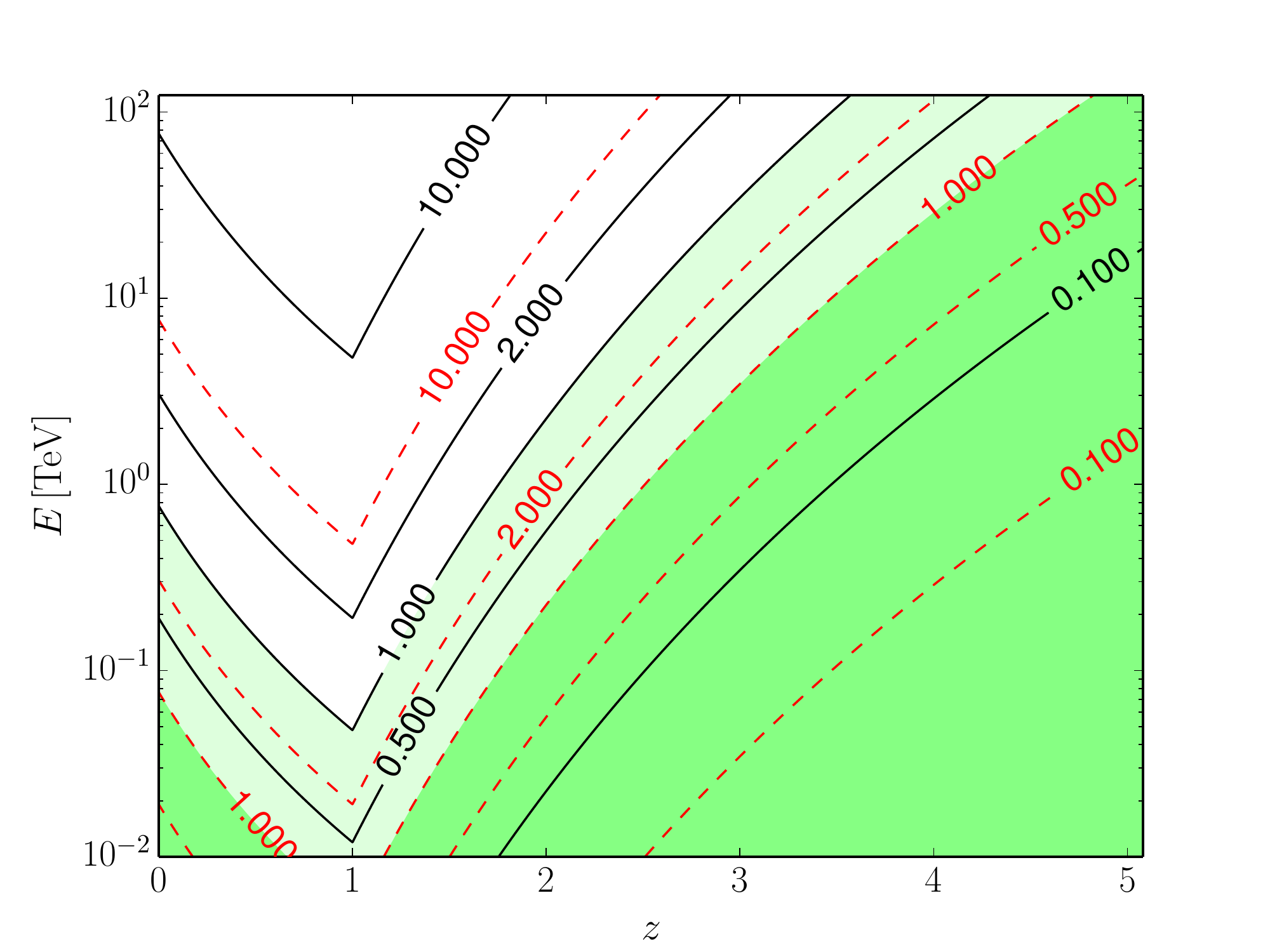}
 \caption{ Contour plot of the ratio between damping due to plasma effects ($\Gamma_{\rm NLD}$) and damping due to inverse Compton scattering ($\Gamma_{\rm IC}$) as a function of redshift and photon energy, $E$, for a blazar with equivalent isotropic luminosity, $EL_E = 10^{45}\,{\rm ergs\,s}^{-1}$ (black solid lines), and $10^{46}\,{\rm ergs\,s}^{-1}$ (red dashed lines).  Lines denoting ratios of $0.1$, 0.5, 1.0, 2.0 and 10.0 are shown.  The shade regions denotes where inverse Compton scattering dominates plasma effects for $EL_E = 10^{45}\,{\rm ergs\,s}^{-1}$ (light green) and $10^{46}\,{\rm ergs\,s}^{-1}$ (dark green). As discussed in the text, this is likely the ``worse case'' scenario for heating due to plasma effects as the inclusion of nonlinear wave-wave damping likely improves the efficiency of the ``oblique'' instability.\label{fig:contour}}
\end{figure}

Figure \ref{fig:contour} shows contours of the ratio between the damping rates due to plasma effects in equation (\ref{eq:nld_damping}) and due to inverse Compton scattering in equation (\ref{eq:ic_damping}).  Lines denoting ratios of 0.1, 0.5, 1.0, 2.0 and 10.0 are shown and the shaded green region denotes where inverse Compton scattering dominates plasma effects. For larger photon energies, the dominance of plasma effects becomes more pronounced.  In addition, the dominance of plasma effects are also more pronounced for increasing redshifts up to $z=1$.  Beyond $z>1$, the constant physical density of the extragalactic background light implies that the stronger $(1+z)^4$ scaling of inverse Compton scattering will become more and more important for the energy budget of these TeV beams.

The limits derived above assume that NLD alone limits the growth rate of the linear ``oblique'' instability.  
We have already noted that this neglects nonlinear wave-wave coupling and quasilinear effects.  However, the modulation instability may also play an important role in limiting the growth of the long-wavelength modes that are fed by NLD. The modulation instability is a result of the ponderomotive force that results from a spatially inhomogeneous distribution of Langmuir waves.  In particular, it can be shown that the change on the time-averaged electron density is $\delta n_e \propto -|E|^2 \propto -W$, i.e., regions of high wave density correspond to regions with lower electron density \citep{Boyd}.  Such a shift in the electron density leads to a shift in the plasma frequency that is of order 
\begin{equation}\label{eq:delta omegap}
\frac{\delta \omega_p}{\omega_p} = \frac{\delta n_e}{2n_e}.
\end{equation}  
As a result sufficiently long-wavelength Langmuir waves that propagate freely in regions of high wave density (low electron density) can lie below the plasma resonance outside, and therefore are trapped.
Which modes become trapped depend upon the shift in the plasma frequency and therefore the wave density.  Assuming the Langmuir wave dispersion relation,
$\omega(k) = \omega_p(1 + 3k^2\lambda_D^2/2$), where $\lambda_D = v_e/\omega_p$ is the Debye length, the condition for trapped modes is given by
\begin{equation}
\frac {W}{n_e k_B T} \gg 3(k\lambda_D)^2.
\label{eq:modulation}
\end{equation} 
The late-time development of the modulation instability in multiple dimensions is currently believed to result in strong turbulence, which then rapidly damps the participating Langmuir waves and results in direct heating of the background plasma (ME13, \citealt{Schlickeiser+12}).

For the long wavelength modes generated by NLD, $k \approx 10^{-2}\omega_p/c$, equation (\ref{eq:modulation}) implies
\begin{equation}\label{eq:modulation2}
\frac {f_w n_b \gamma_{b} m_e c^2}{n_e k_B T} \gg \left(\frac{3\times 10^{-2}v_e}{c}\right)^2 \approx 6\times 10^{-10}\left(\frac T {10^4\,{\rm K}}\right)
\end{equation}
where $f_w$ is the saturation amplitude of these waves relative to the beam energy $n_b \gamma_b m_e c^2$.  For $f_w = 10^{-3}$ given by the saturation amplitude due to NLD, this gives 
\begin{equation}
\frac {f_w n_b \gamma_{b} m_e c^2}{n_e k_B T}\approx  2\times 10^{-9}\left(\frac{f_w}{10^{-3}}\right)\left(\frac {n_b/n_e} {3\times 10^{-18}}\right)\left(\frac {\gamma_{b}} {10^6}\right)\left(\frac T {10^4\,{\rm K}}\right)^{-1}, 
\end{equation}
which is above the criterion given in equation (\ref{eq:modulation2}) and can allow these waves to rapidly heat the background electrons \citep{Schlickeiser+12}.  In doing so, the modulation instability may limit the effectiveness with which NLD can drive  long-wavelength modes, and hence the linear growth of the instability, as well as provide a natural mechanism for converting the wave energy into heat. However, more work is required on this question.

\section{Conclusions}\label{sec:conclusions}

We have calculated the effects of NLD on the saturation amplitude of the ``oblique'' instability to which high energy pair beams in the IGM are linearly unstable to.  Using a numerical calculation, we find that the ``oblique'' instability remains the most powerful cooling mechanism for these pair beams contrary to the earlier claims of ME13.  In particular, we find that the beam  saturates at a rate that is $\approx 0.1\%$ of kinetic growth rate of the ``oblique'' instability.  The damping of the beam leads to the transfer of beam energy into long-wavelength non-resonant waves.  When comparing to the estimate of ME13, we find that our damping rate exceeds their estimate by two orders of magnitude.  Using this damping rate, we conclude that the oblique instability is effective in quenching the beam at $z \approx 1$, but it less effective at different redshifts. 

We caution that the results that we present here are limited to the effects of NLD.  As we argue above, the inclusion of nonlinear wave-wave coupling will likely lead to a more equitable distribution of energy among wavevectors.  The nonlinear damping rate of the ``oblique'' instability is likely to increase under these conditions.  Thus, the calculation that we present here represents the ``worst-case'' scenario for the ``oblique'' instability where the magnitude of plasma effects is comparable to the effects of inverse Compton scattering.

\acknowledgments P.C. and A.L. gratefully
acknowledges support from the NASA ATP
program through NASA grant NNX13AH43G, and the NSF through grant AST-1255469. 
A.E.B.~and M.S.~receive financial support from the Perimeter
Institute for Theoretical Physics and the Natural Sciences and
Engineering Research Council of Canada through a Discovery Grant.
Research at Perimeter Institute is supported by the Government of
Canada through Industry Canada and by the Province of Ontario through
the Ministry of Research and Innovation.
C.P.~gratefully acknowledges
financial support of the Klaus Tschira Foundation. E.P. acknowledges support by the ERC grant ``The Emergence of Structure during the epoch of Reionization''.

\bibliographystyle{apj}

\end{document}